# Upper critical magnetic field in $Ba_{0.68}K_{0.32}Fe_2As_2$ and $Ba(Fe_{0.93}Co_{0.07})_2As_2$

V.A. Gasparov[1], L. Drigo[2], A. Audouard[2], D.L. Sun[3], C.T. Lin[3], S. L. Bud'ko[4], P. C. Canfield[4], F. Wolff-Fabris[5] and J. Wosnitza[5]

[1] *Institute of Solid State Physics RAS, Chernogolovka, 142432, Russian Federation*
[2] *Laboratoire National des Champs Magnétiques Intenses (UPR 3228 CNRS, INSA, UJF, UPS), Toulouse, France*
[3] *Max Planck Institute for Solid State Research, 70569 Stuttgart, Germany*
[4] *Ames Laboratory, US DOE and Department of Physics and Astronomy, Iowa State University, Ames, Iowa 50011, USA*
[5] *Hochfeld-Magnetlabor Dresden (HLD), Helmholtz-Zentrum Dresden-Rossendorf, 01314 Dresden, Germany*

**Abstract**

We report measurements of the temperature dependence of the radio-frequency magnetic penetration depth in $Ba_{0.68}K_{0.32}Fe_2As_2$ and $Ba(Fe_{0.93}Co_{0.07})_2As_2$ single crystals in pulsed magnetic fields up to 60 T. From our data, we construct an $H$-$T$ phase diagram for the inter-plane ($H \parallel c$) and in-plane ($H \parallel ab$) directions for both compounds. For both field orientations in $Ba_{0.68}K_{0.32}Fe_2As_2$, we find a concave curvature of the $H_{c2}(T)$ lines with decreasing anisotropy and saturation towards lower temperature. Taking into account Pauli spin paramagnetism we can describe $H_{c2}(T)$ and its anisotropy. In contrast, we find that Pauli paramagnetic pair breaking is not essential for $Ba(Fe_{0.93}Co_{0.07})_2As_2$. For this electron-doped compound, the data support a $H_{c2}(T)$ dependence that can be described by the Werthamer-Helfand-Hohenberg model for $H \parallel ab$ and a two-gap behavior for $H \parallel c$.

**Introduction**

Although the Fermi surfaces of the iron pnictides are definitely quasi-two-dimensional [1], reports on the anisotropy of the upper critical field, $H_{c2}(T)$, are quite puzzling [1-11]. In the field range below 10 T, where $H_{c2}(T)$ is limited by orbital pair breaking, $H_{c2}(0)$ is evaluated through the slope of $dH_{c2}/dT|_{Tc}$ close to $T_c$ according to the well-known Werthamer–Helfand–Hohenberg (WHH) model for the orbital critical magnetic field $H^{orb}_{c2}(0) \approx -0.69 T_c (dH_{c2}/dT)|_{Tc}$ [12]. A significant anisotropy of $\gamma = H^{ab}_{c2}(0)/H^{c}_{c2}(0)$ is reported for $NdFeAsO_{0.9}F_{0.1}$ [6], NaFeAs with Co and P doping [7], and $Ba(Fe_{0.93}Co_{0.07})_2As_2$ (hereafter BFC) in this field range [9]. However, direct measurements of $H_{c2}(T)$ in pulsed magnetic fields have shown that the actual anisotropy of $H_{c2}(0)$ becomes very small at low temperatures [3,4,9,10]. Recently, we have shown by use of resistivity measurements that in $Ba_{0.68}K_{0.32}Fe_2As_2$ (hereafter BKF) this anisotropy is washed out in strong fields due to Pauli spin paramagnetism [11].

Here, we complement the study of the upper critical fields parallel and perpendicular to the crystallographic $c$ axis in a high-quality hole-doped BKF single crystal by aid of a radio-frequency tunnel-diode-oscillator technique. In addition, we present as well data for an electron-doped BFC single crystal. We show that for BKF the whole $H_{c2}(T)$ dependence can be explained taking into account Pauli spin paramagnetism. The latter substantially limits $H_{c2}(T)$ and, correspondingly, the anisotropy at lower temperatures. In contrast, we find that Pauli paramagnetic pair breaking is not essential for BFC.

**Experimental**

Optimally doped BKF single crystals were grown from FeAs formed by selflux in a zirconia crucible sealed in a quartz ampoule under argon atmosphere. Single crystals of BFC were grown out of FeAs flux using a high-temperature solution growth techniques. More details on the growth methods, crystal structures, and characterization are given elsewhere [8,11,13,14,15]. The samples were plates with dimensions of about 1x1x0.15 mm$^3$ for BKF and 1x1x0.2 mm$^3$ for BFC. Bulk superconductivity was confirmed by magnetic-susceptibility and *dc*-conductivity measurements. The resistive superconducting transition temperatures for the studied samples are $T_c$ = 38.5 and 22 K, for BKF and BFC, respectively. The BKF single crystal has the lowest residual specific heat reported so far for FeAs-based superconductors. The ratio of the resistivities at 300 K with respect to the value just above $T_c$ is 16 [11] and the ratio of the residual electronic specific heat with respect to the normal-state value yields an estimate for the non-superconducting fraction of less than 2.4 % [13]. These features show the strongly reduced residual impurity scattering and hence high sample quality and purity.

The device for the radio frequency (RF) magnetic penetration depth measurements is a LC-tank circuit powered by a tunnel diode oscillator (TDO) biased in the negative resistance region of the current voltage characteristic, as reported in Ref. [16]. A mica-chip capacitor connected by a semi-rigid 50 Ω coaxial cable to a pair of counter-wound coils is used. The coils are made out of copper wire (100 μm in diameter) wound around a Kapton tube with a diameter of 1.1 mm. A compensated form of the RF coil was necessary to minimize induced voltages during the field pulse. After signal amplification, mixing with a reference signal and demodulation, the resulting oscillator frequency, which can be approximated by $f = 1/(2\pi\sqrt{LC})$, lies in the MHz range. The fundamental resonant frequency is approximately 1 MHz at $T_c$.

The RF technique was used because it provides a contactless measurement, much more sensitive than conventional four-point technique for low-resistance samples such as superconductors at low temperatures [2]. The samples were placed inside the counter-wound coil pair parallel or perpendicular to the coil axis. This latter configuration has a smaller filling factor than for the parallel configuration resulting in a smaller but still easily resolvable frequency shift on a large background. As the magnetic field increases, the transition to the normal state is detected from the shift in the resonance frequency. The resulting frequency variation versus magnetic field is, in first order, proportional to changes in the magnetic penetration depth. The temperature was stabilized by use of a Lake Shore temperature controller with an accuracy of ±0.1 K.

The experiments were performed at fixed temperatures in pulsed magnetic fields of up to 60 T, with pulse-decay durations of 250 ms, at the Laboratoire National des Champs Magnétiques Intenses of Toulouse (CNRS). The magnetic field was applied either along the *c* axis or in the *ab* plane. Even though the reported data are collected during the decaying part of the pulse, we have checked that they are in agreement with data taken at the rising part, although with a reduced signal-to-noise ratio in the latter case, which confirms that the data are not affected by sample heating during the pulse.

**Results and discussion**

Figure 1 shows the temperature dependences of the TDO frequency changes and the resistances of BKF (a) and BFC (b) at zero magnetic field. Magnetic-susceptibility data at small fields (10 Oe) are shown in Fig. 1(a) as well. Very sharp superconducting transitions are observed in all quantities. From the middle point in the resistive transition, $T_c \approx$ 38.5 K for BKF is obtained. Above $T_c$, an almost linear temperature dependence of the resistivity, $\rho(T)$, with a tendency to saturation at high temperatures is found. Recently, we have shown that the whole $\rho(T)$ dependence is well described by an exponential term due to intersheet electron-phonon umklapp scattering between light electrons around the *M* point and heavy hole sheets at the *Γ* point in reciprocal space [11]. The TDO data (Fig. 1) yield $T_c$ values somewhat lower than the resistivity and magnetic-susceptibility data. This feature is discussed below in comparison with $H_{c2}(T)$ data derived from *R(H)* [11]. Figures 2 and 3 display the magnetic field dependences of the TDO frequency in pulse magnetic fields up to 60 T aligned parallel [Fig. 2(a) and Fig. 3(a)] and perpendicular [Fig. 2(b) and Fig. 3(b)] to the *c* axis for BKF and BFC, respectively. In line with the large $T_c$ values reported in Fig. 1, the materials show superconducting transitions up to very high fields.

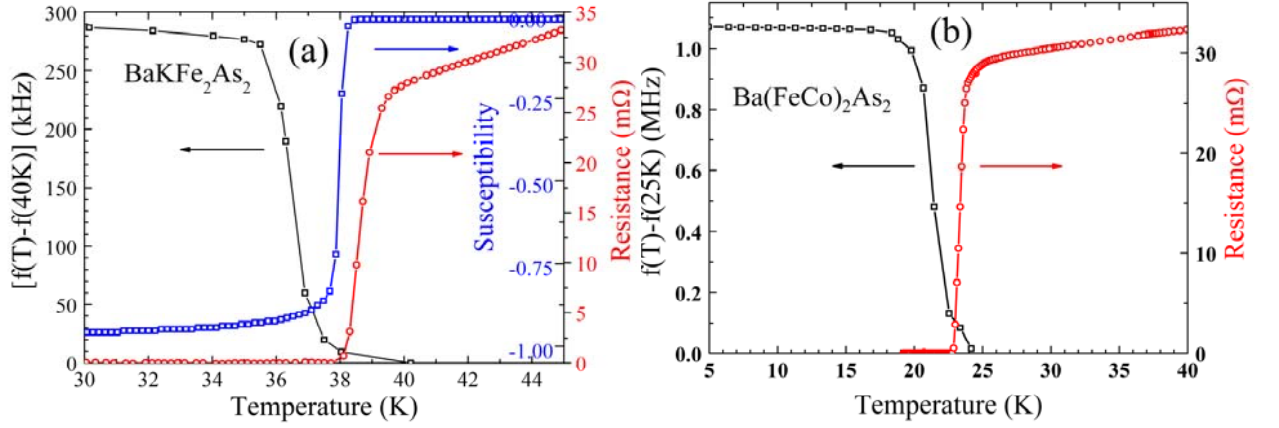

Fig. 1 (Color online) Temperature dependences of the TDO frequency, resistance and susceptibility for (a) BKF and (b) BFC.

The superconducting transitions in applied fields are not substantially broadened and are much narrower than reported e.g. for $Ba_{0.55}K_{0.45}Fe_2As_2$ [2]. The transition curves just move to higher fields with decreasing temperature for both field orientations. Compared to resistivity data for BKF [11], the RF curves are narrower than the resistive transitions. This can be understood keeping in mind that RF measurements are not dependent on a macroscopic net current flow across the sample [2]. The method for determining consistent $H_{c2}$ values from the data shown in Figs. 2 and 3 is based on identifying the point at which the steepest slope of the RF signal at the transition intercepts with the linearly extrapolated normal-state background as discussed in Ref. [2] [see the construction lines in Figs. 2(a) and 3(a)]. As seen from Figs. 2(a) and 3(a), the high-field normal-state background data at high *T* exhibit a close to linear dependence of $\Delta f(H)$ for $H \parallel c$ and a curved one for $H \parallel ab$ [Figs. 2(b) and 3(b)]. Due to the smaller filling factor, a large background is observed in the normal state for $H \parallel ab$. Approximating the background by a polynomial, the superconducting part of the signal can be extracted from the data. As an example, results for BKF are displayed in the inset of Fig. 2(b). The same data treatment was done for BFC for $H \parallel ab$ as well.

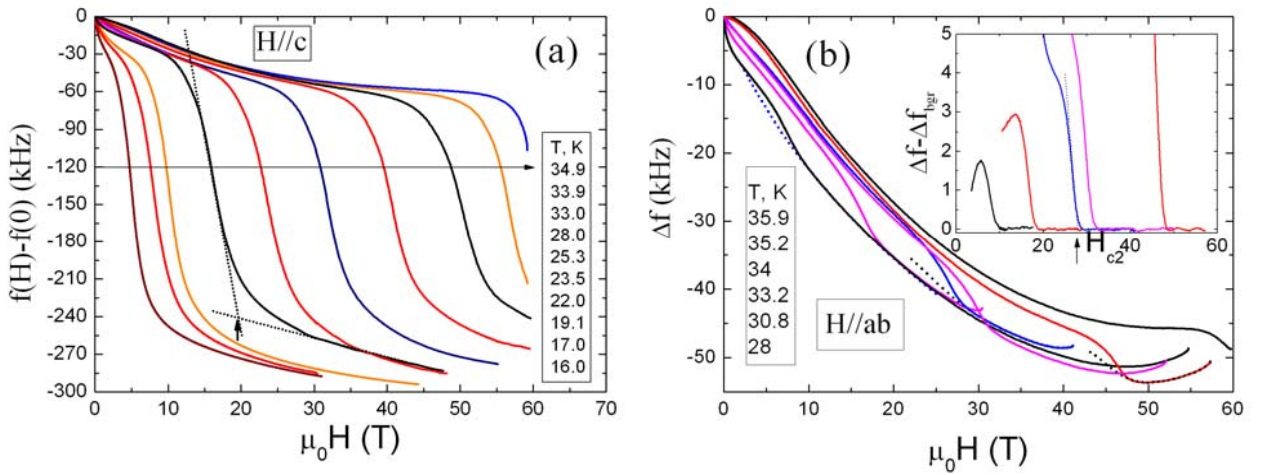

Fig. 2 (Color online) Field dependences of the TDO frequency shifts for BKF for magnetic fields applied (a) along the *c* direction and (b) parallel to the *ab* plane at temperatures from 16 to 36 K.

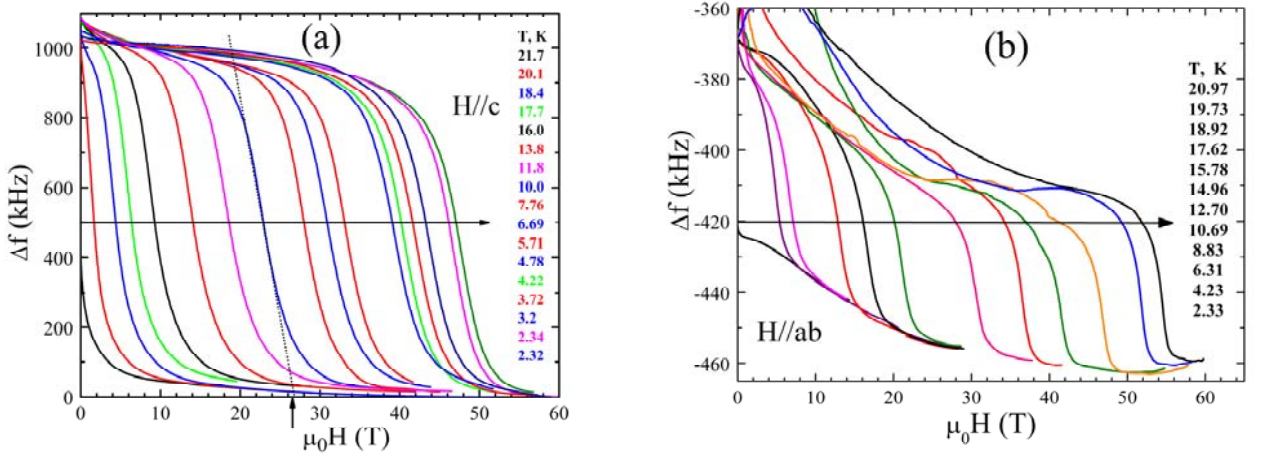

Fig. 3 (Color online) Field dependences of the TDO frequency shifts for BFC for magnetic fields applied (a) along the *c* direction and (b) parallel to the *ab* plane at temperatures from 2 to 22 K.

The resulting temperature dependences of $H_{c2}^{ab}$ and $H_{c2}^{c}$ for the almost optimally hole-doped BKF and electron-doped BFC samples are shown in Figs. 4(a) and 4(b), respectively. The data obtained earlier by use of resistivity measurements [11] are plotted as red squares. As mentioned above, TDO data yield $T_c$ values lower than those deduced from resistivity measurements (see Fig. 1). For this reason, normalized temperatures $T/T_c$, are plotted in Fig. 4a. Close to $T_c$, the usual linear temperature dependence of $H_{c2}$ is observed, with clearly different slopes for the two field orientations. Towards lower temperatures, a clear saturation for both field orientations for BKF is observed. The anisotropy parameter $\gamma$, which is about 2.0 near $T_c$, decreases considerably at low temperature (see the insets Fig.4a and 4b).

Apparently, the small anisotropy of $H_{c2}(T)$ is due to a partial compensation of the orbital pair-breaking mechanism by Pauli paramagnetism, rather than due to Fermi-surface effects. The temperature-dependent anisotropy $\gamma$ we observed is most likely due to two independent pair-breaking mechanisms [11,17]: (i) at higher temperatures, Cooper pairing is suppressed by orbital currents that screen the external field; (ii) towards lower temperatures, the limiting effect is caused by the Zeeman splitting, i.e., when the Zeeman energy becomes larger than the condensation energy the Pauli limit, $H_p$, is reached [18,19]. In a simple approximation assuming that the superconducting gap, $\Delta$, is given by $2\Delta = 3.5 k_B T_c$, $H_p$ is *1.84 $T_c$ [T/K]* [18], resulting in $H_p = 71$ *T* for BKF. This paramagnetic limit is lower than the orbital limit, $H^*_{c2}(T)$, which is related to the slope of $H_{c2}(T)$ close to $T_c$. With the experimental slopes for $dH^c_{c2}/dT = -4.76$ T/K and $dH^{ab}_{c2}/dT = -9.93$ T/K for BKF for $H \parallel c$ and $H \parallel ab$, respectively, the WHH model [12] predicts $H^{*c}_{c2}(0) = 120\ T$ and $H^{*ab}_{c2}(0) = 250\ T$ at $T = 0$, respectively. For BFC, $dH^c_{c2}/dT = -2.42$ T/K and $dH^{ab}_{c2}/dT = -4.0$ T/K result in much lower estimates: $H^{*c}_{c2}(0) = 35\ T$ and $H^{*ab}_{c2}(0) = 58\ T$ at $T = 0$. The dashed and dotted lines in Figs. 4(a) and 4(b) display the temperature dependence of the orbital critical fields within the WHH approach for both field orientations and compounds ignoring the Pauli limit. These $H^*_{c2}(0)$ values, allow to derive the coherence lengths $\xi(0)$. We obtain $\xi^{ab}(0) = \sqrt{\phi_0 / 2\pi H^{*c}_{c2}} = 1.32$ *nm* and $\xi^c(0) = \phi_0 / 2\pi \xi^{ab} H^{*ab}_{c2} = 0.64$ nm for BKF, and $\xi^{ab}(0) = 2.45\ nm$ and $\xi^c(0) = 1.48\ nm$ for BFC, respectively. Although anisotropic, the *c*-axis coherence lengths for $Ba_{0.68}K_{0.32}Fe_2As_2$, is nevertheless larger than the thickness of 0.32 nm of the conducting FeAs sheet indicating the three-dimensional nature of the superconductivity for both compounds. When including Pauli paramagnetism, the upper critical field is reduced relative to $H^*_{c2}(T)$ to [17,18,19]:

$$H_{c2}(T) = H^*_{c2}(T)/[1 + \alpha^2(T)]^{1/2}, \qquad (2)$$

where $\alpha(T) = \sqrt{2} H^*_{c2}/H_p$ is the Maki parameter. The solid lines in Fig. 4a are the best fits using Eq. (2). A very good agreement with the experimental data is observed for both field orientations for BKF, with only one free parameter, namely $H_p = 134\ T$. This value is twice as large as the above estimate of 71 T. Nevertheless, this discrepancy is not unexpected since in the latter value neither many-body correlations nor strong-coupling effects are included [17]. Actually, an estimate of $H_p$ by use of ARPES data for the superconducting gap, $2\Delta(0) = 7.7 k_B T_c$ [20,21], and using the Clogston equation $H_p = \Delta(0)/\sqrt{2}\mu_B$ [18] results in $150\ T$, which is in good agreement with the Pauli limit of $134\ T$ extracted from our data.

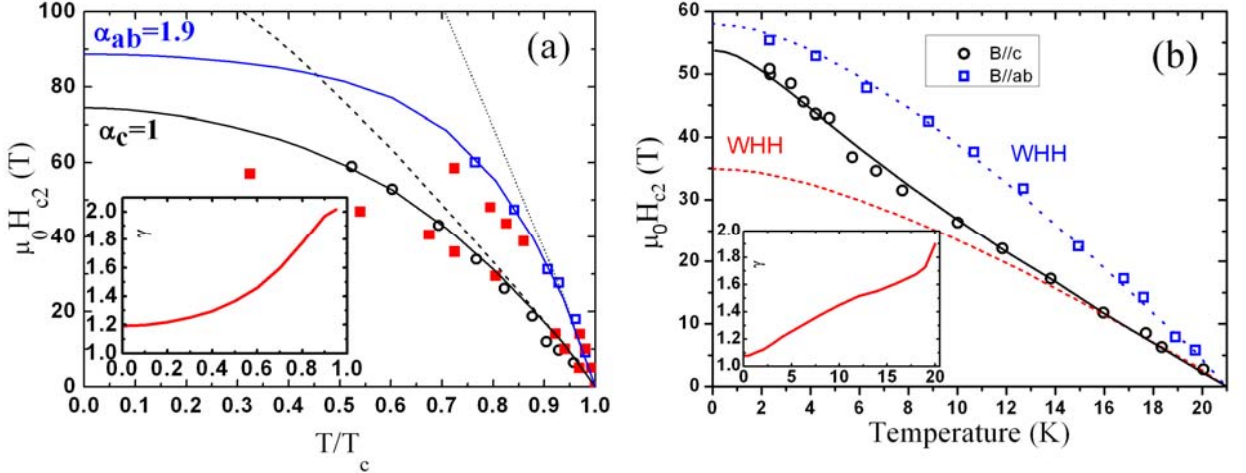

Fig. 4 (Color online) Temperature dependences of the upper critical fields, $H_{c2}(T)$, for (a) BKF and (b) BFC for $H \parallel c$ (black circles) and $H \parallel ab$ (blue squares). Data from transport measurements for BKF are shown by red solid squares [11]. The dotted and dashed lines indicate the temperature dependences according to the WHH model neglecting Pauli limiting for both samples. The solid lines in (a) show the dependences including Pauli pair breaking using Eq. (1) with only one free parameter, $H_p=134\ T$, for both field orientations. The solid curve in (b) is a two-band fit by use of Eq.(3), while the blue dashed curve for $H \parallel c$ is calculated by use of the WHH model. Inset: temperature-dependent anisotropy of $\gamma = H^{ab}_{c2}/H^{c}_{c2}$ as determined from the fits to the $H_{c2}(T)$ data.

A somewhat different behavior is observed for underdoped $Ba_{0.55}K_{0.45}Fe_2As_2$ ($T_c = 32\ K$) [2] and $Ba_{1-x}K_xFe_2As_2$ with $x \approx 0.4$ ($T_c = 28\ K$) [3]. In these studies, $H_{c2}$ goes to saturation both for $H \parallel ab$ and $H \parallel c$. We reanalyzed the data from Ref. [2] keeping in mind Pauli paramagnetsim described above. Although the data for $H \parallel ab$ are in good agreement with our results, with $H_p = 114\ T$, $H_{c2}(T)$ follows the WHH temperature dependence down to 0 K for $H \parallel c$. This most probably is due to an orbital critical field smaller than the Pauli limit for $Ba_{0.55}K_{0.45}Fe_2As_2$ for this field orientation. Very small anisotropies $\gamma$ are observed in Refs. [2,3,9]. Oppositely, the anisotropy values of oxifluoride Fe - 1111 compounds are substantially larger than for the 122 materials, and a transition to the normal state at low temperatures is not achieved even in pulsed fields above 60 T [3,7].

However, as shown in Fig. 4(b), the Pauli scenario cannot account for the data of electron-doped BFC for $H \parallel c$. The simple estimate, $H_p = 71.8\ T$, using the Clogston equation and ARPES data for the superconducting gap $2\Delta(0) = 6.5\ k_B T_c$ (for the $\beta$ hole barrel of the Fermi surface) [22,23] is two times larger than $H^{*c}_{c2}(0) = 35\ T$, indicating that Pauli paramagnetic pair breaking is not essential for BFC. Even though $H^{*ab}_{c2}(0) = 58\ T$ is substantially larger, it is still lower than $H_p$. Indeed, as can be seen in Fig. 4(b), the WHH model nicely fits the data for $H \parallel ab$. For $H \parallel c$, however, we observe a positive curvature at low temperatures without saturation. Apparently, the upward curvature of $H^{c}_{c2}(T)$ originates from two-band features recently evidenced by resistive data for BFC [9]. According to Gurevich [24], the zero-temperature value of the $H_{c2}$ is significantly enhanced in the two gap dirty limit superconductor model,

$$H_{c2}(0) = \frac{\phi_0 k_B T_c}{1.12\hbar\sqrt{D_1 D_2}} \exp(\frac{g}{2}) \quad , \quad (2)$$

as compared to the one-gap dirty-limit approximation $H_{c2}(0) = \phi_0 k_B T_c/1.12\hbar D$. Here, $g$ is a rather complicated function of the matrix of the BCS superconducting coupling constants $\lambda_{mm'} = \lambda^{(ep)}_{mm'} - \mu_{mm'}$, where $\lambda^{(ep)}_{mm'}$ are electron-phonon coupling constants and $\mu_{mm'}$ is the matrix of the Coulomb pseudopotential. In a simple approximation using the same inter-band, $\lambda_{12} = \lambda_{21} = 0.5$, and intra-band, $\lambda_{22} = \lambda_{11} = 0.5$, coupling constants [9], the equation for $H_{c2}(T)$ takes the simple Usadel form:

$$a_1[\ln t + U(h)] + a_2[\ln t + U(\eta h)] = 0. \quad (3)$$

Here, $a_1 = 1 + \lambda_-/\lambda_0 = 1$; $a_2 = 1 - \lambda_-/\lambda_0 = 1$; $\lambda_0 = (\lambda_-^2 + 4\lambda_{12}\lambda_{21})^{1/2} = 1$; $\lambda_- = \lambda_{11} - \lambda_{22} = 0$; $h = H_{c2} D_1 \hbar/2\phi_0 k_B T$; $\eta = D_2/D_1$; $U(x) = \Psi(1/2 + x) - \Psi(1/2)$, where $\Psi(x)$ is the digamma function, $t = T/T_c$, $\phi_0$ is the magnetic flux quantum, and $D_{1,2}$ are the electronic diffusivities for different Fermi-surface sheets [24]. We assume that the derivative $dH_{c2}/dT = 2.37$ T/K close to $T_c$ is determined by $D_1$ for the band with the highest coupling constant, i.e., $D_1 \gg D_2$ [25], and thus estimate $D_1$ from:

$$D_1 \approx \frac{8\phi_0 k_B}{\pi^2 \hbar dH_{c2}/dT} = 0.926 \text{ cm}^2/\text{sec}. \quad (4)$$

Given this $D_1$, the temperature dependence of $H_{c2}(T)$ is accounted for by Eq. (3) with $\eta=0.12$ [solid curve in Fig. 4(b)]. Therefore, the limiting value of $H_{c2}(0)$ is likely dominated by a band with low diffusivity $D_2 = 0.111$ cm$^2$/sec, while the slope $dH_{c2}/dT$ close to $T_c$ is due to a band with larger diffusivity, $D_1$. This two-gap model quantitatively reproduces the unconventional non-WHH temperature dependence of $H_{c2}$ for $H \parallel c$, while the one-gap WHH model works nicely for $H \parallel ab$. The overall dependence with $\eta=0.12$, is in agreement with earlier data for BFC at $H \parallel c$ with slightly lower $\eta=0.085$ [9]. However, in contrast to Ref. [9], we did not find any significant influence of Pauli paramagnetism for $H \parallel ab$, probably because $H_p$ is too large.

We can now turn to one important point of this paper which is the comparison of the electronic structure of BFC and BKF in order to better understand the features observed. Angular-resolved photoemission spectroscopy (ARPES) data [20,21] of Ba$_{1-x}$K$_x$Fe$_2$As$_2$ reveal four Fermi surface (FS) sheets: two concentric corrugated cylindrical hole barrels around the $\Gamma$ point (with Fermi wave vectors $k_{F\alpha} \approx 0.2\pi/a$ for the $\alpha$ sheet and $k_{F\beta} \approx 0.45\pi/a$ for the $\beta$ sheet) and two electron tubes $\gamma$ and $\delta$ around the $M$ points (with almost identical $k_F \approx 0.2\pi/a$). The averaged SC gaps on each Fermi surface (FS) is nearly isotropic whereas the gap values are strongly different: 12.3, 5.8, 12.2, and 11.4 meV for the $\alpha$, $\beta$, $\gamma$, and $\delta$ FS sheets, respectively [21]. The estimates for the effective masses of these bands obtained from ARPES data are $m^*_\alpha = 4.8$, $m^*_\beta = 9.0$, $m^*_\gamma = m^*_\delta = 1.3$ in units of the free-electron mass $m_0$ leading to rather isotropic Fermi velocities of 0.5, 0.22, 0.32, 0.48 eVÅ for these bands, respectively [22].

According to studies for electron-doped BFC [23,24], the FSs consist of only one hole-like FS sheet, $\beta$, centered at the $\Gamma$ point and two electron-like FS sheets, an inner ($\gamma$) and outer ($\delta$) pocket near the $M$ point. The latter is as for the hole-doped BKF. The FS volume of the $\beta$ pocket and the ellipsoidal electron pockets was estimated to be 1.6% and 3.2% of the unfolded first Brillouin zone, respectively. The smaller but isotropic superconducting gap $\Delta(0)$ values of the $\gamma$ and $\delta$ FS sheets (5 meV) are close to that of the hole-like $\beta$ sheet (6.5 meV). The effective masses of these bands are not known for BFC, but we may assume that they are similar as those for Ba$_{1-x}$K$_x$Fe$_2$As$_2$.

Thus, in both materials similar superconducting gaps are found. Apparently, the difference observed in $H_{c2}(T)$ might be due to the larger electron sheets in BFC compared to BKF and the different $T_c$'s. Although the two gap features observed for BFC can be modeled by two independent BCS superconducting bands with different plasma frequencies, gaps, and $T_c$'s [26], it is not clear why this model works for $H \parallel c$ only. For $H \parallel ab$, this material behaves as a single-band BCS superconductor.

*In conclusion*, the measurements of $H_{c2}(T)$ for an electron- and a hole-doped *122* iron-pnictide superconductors let us conclude that: (i) for hole-doped BKF, we can account for the temperature dependence of $H_{c2}$ with only one fitting parameter, namely, the Pauli-limiting field, over the whole temperature region for both $H \parallel c$ and $H \parallel ab$, (ii) for electron-doped BFC, the data support a conventional WHH $H_{c2}(T)$ dependence for $H \parallel ab$ and a two-gap behavior for $H \parallel c$ indicating that Pauli paramagnetic pair breaking is not essential for this compound, (iii) the ratio of the diffusivities for the two-band model is rather large, $D_1/D_2 = 8.4$, indicating that the scattering rates of the bands differ by about one order of magnitude, (iv) the coherence length is anisotropic in both compounds and is larger than the thickness of the conducting sheets indicating 3D superconductivity.

**Acknowledgment**

We would like to thank V.F. Gantmakher and R. Huguenin for helpful discussions and to G. Ballon for help during experiments. Part of this work has been supported by EuroMagNET II under the EU contract 228043 and FP7 I3, the German Science Foundation within SPP 1458, and RAS Program: New Materials and Structures (Grant B.7) and the Russian State Program of support of leading scientific schools (1160.2008.2). Work done by PCC and SLB was supported by the U.S. Department of Energy, Office of Basic Energy Science, Division of Materials Sciences and Engineering. The research was performed at the Ames Laboratory which is operated for the U.S. Department of Energy by Iowa State University under Contract No. DE-AC02-07CH11358.